\documentclass[smallextended]{svjour3}       
\usepackage{natbib}
\usepackage{graphicx}
\usepackage{url}
\usepackage{tabularx}
\usepackage{color}
\usepackage{comment}
\usepackage[flushleft]{threeparttable}

\journalname{}
\date{}

\begin{document}

\title{Workflow Management on the Blockchain --- Implications and Recommendations}
\titlerunning{Blockchain WFMS}

\author{Joerg Evermann
\and
Henry Kim}

\authorrunning{J. Evermann and H. Kim}

\institute{J. Evermann \at Memorial University of Newfoundland, St. John's, Canada\\
\email{jevermann@mun.ca}\\
\url{https://joerg.evermann.ca} \\ \and
H. Kim \at York University, Toronto, Canada}

\maketitle

\begin{abstract}
Blockchain technology, originally popularized by cryptocurrencies, has been proposed as an infrastructure technology with applications in many areas of business management. Blockchains provide an immutable record of transactions, which makes them useful in situations where business actors may not fully trust each other. The distributed nature of blockchains makes them particularly suitable for inter-organizational e-Business applications. In this paper we examine the use of blockchains for executing inter-organizational workflows. We discuss architectural options and describe prototype implementations of blockchain-based workflow management systems (WfMS), highlighting differences to traditional WfMS. Our main contribution is the identification of potential problems raised by blockchain infrastructure and recommendations to address them.

\end{abstract}

\section{Introduction}

Workflow management (WfM) has traditionally focused on intra-enterprise coordination of work items and resources. Despite its many challenges, inter-organizational WfM has seen less research attention. In particular, inter-or\-ga\-ni\-zation\-al processes may include stakeholders that are in adversarial relationships with each other, but that nonetheless have to jointly complete process instances. In such situations, trust in the current state of a process instance and the correct execution of workflow activities may be lacking. Blockchain technology can help by providing a trusted, distributed, workflow execution infrastructure.

A blockchain cryptographically signs a series of blocks that contain transactions, providing an immutable record. In a distributed blockchain, actors form a peer-to-peer (P2P) network to independently validate transactions and add them to the block chain. In inter-organizational workflow management, actors must agree on the state of work as this determines the set of next valid activities in the process. Thus, it is natural to use blockchain transactions to record either workflow activities or workflow states. 

Workflow management systems (WfMS) can be implemented in different ways on blockchain infrastructure. In contrast to earlier work, we do not use smart contracts to implement model-specific workflow engines. \emph{While smart contracts have great potential for workflow management, we do not believe that this is necessarily the only way to integrate blockchain technology with WfMS. Further, we believe that to conclude that the problem is solved is premature at this time, given that the issue has only very recently received research attention \cite{mendling2018blockchains}, and as witnessed by some of the issues we identify in this paper. Moreover, given the extensive investment in WfMS by researchers and practitioners, we believe that investigating how existing, ''standard'' WfMS can be implemented on blockchain infrastructure \emph{without re-implementation} in smart contract languages is worthwhile. To our knowledge, there has been no implementation of such a system.} 

In this paper, we show that generic or existing workflow engines can be readily adapted to fit onto a blockchain infrastructure and that smart contracts are not required. To this effect, we investigate and propose standard interfaces between blockchain infrastructure and workflow engines. We describe two research prototype WfMS that provide proof-of-concept implementations\footnote{Source code available from the corresponding author's website at \url{https://joerg.evermann.ca/software.html}} of our proposed architecture. The distributed nature of a blockchain and the nature of the consensus finding process raises challenges that are not seen in centralized WfMS. Our research prototypes have helped us identify differences between traditional WfMS and blockchain-based WfMS. We explore these differences and their implications for workflow execution and offer recommendations for future blockchain-based WfM.

\paragraph*{Contribution:\ }\ While the prototype implementations are important demonstrations of feasibility, they are research tools only. Instead, our main contribution is in the lessons learned from their implementation and our recommendations for blockchain-based WfMS. Specifically, in this paper we:
\begin{enumerate}
\item demonstrate the feasibility of an alternative to smart contracts for block\-chain-based WfMS, including the development of generic interfaces between architecture components, and
\item identify problems, implications, and recommendations arising from the use proof-of-work blockchains for workflow management. These are the same for smart contract-based architectures and for architectures not based on smart contracts, as they stem from the properties of the proof-of-work consensus mechanism.
\end{enumerate}

The remainder of the paper is structured as follows. Section~\ref{sec:related} reviews related work on distributed, inter-organizational, and blockchain-based workflow management. We then briefly describe the main ideas behind distributed blockchains (Sec.~\ref{sec:blockchains}). Following this, Sec.~\ref{sec:principles} presents the main principles of our approach and discusses validity guarantees. Next, Sec.~\ref{sec:prototype} presents our prototype implementations. Implications of using blockchain technology for workflow execution are discussed at length in Sec.~\ref{sec:issues}. We conclude with recommendations for addressing potential problems, a comparison of architectures, and an outlook to future work (Sec.~\ref{sec:discussion}).

\section{Related Work}
\label{sec:related}

A blockchain-based WfMS can be viewed as a type of distributed, replicated, inter-organizational WfMS. Hence, this section reviews prior research on distributed and replicated WfMS, inter-organizational WfMS, and the state-of-the art in blockchain-based WfMS. 

\subsection{Distributed Workflow Management}
\label{sec:dwfms}

Distributed WfMS have seen research interest in the late 1990s and early 2000s. With the advent of client-server technology, distributed object-oriented standards such as CORBA, and the beginnings of P2P networking, researchers identified ways to use these infrastructure technologies to address technical issues such as fault tolerance, redundancy, and scalability through distribution and replication. 

The Exotica/FlowMark system by IBM \citep{alonso1995exotica} focuses on persistent message passing between nodes when the process can be partitioned onto different workflow nodes. In the Ready system \citep{eder1999towards}, independent WfMS can subscribe to a shared event-publishing system. METEOR2 coordinates independent workflow systems using distributed workflow schedulers \citep{das1997orbwork,miller1998webwork}. Workflow evolution in distributed systems has been studied in the ADEPT system \citep{reichert2003adept,reichert2007supporting} while P2P network technology has been used to implement distributed ''web workflow peers'' that execute workflows controlled by a central administration peer \citep{fakas2004peer}. The SwinDeW system \citep{yan2006swindew} is another approach based on P2P technology. Efficiency of network communication has been the focus of \citet{bauer1997distributed}, who develop optimal algorithms for case transfer of sub-workflows to distributed servers. A load-balancing approach by \citet{jin2001load} uses a central decision making component to distribute complete workflow instances across multiple WfMS. The event-based distributed system EVE \citep{geppert1998event} relies on synchronized clocks to distribute workflow activities to participating service execution nodes. Based on partitioning of state-charts and incremental synchronization of distributed workflow engines, the Mentor project \citep{muth1998centralized} developed algorithms for optimal communication and message exchange among distributed WfMS. The Metuflow system \citep{dogac1998design} uses transaction semantics to determine the proper sequence of activities in a distributed system that is built on a reliable message passing infrastructure and CORBA message exchange. CORBA also forms the infrastructure for an approach that uses a common monitor and scheduler to coordinate multiple ''task managers'' that can independently execute workflow activities \citep{miller1996corba}. The Wasa2 system \citep{vossen1999wasa2} also implements a CORBA-based infrastructure of services to manage business and workflow objects. Focusing on performance and availability, continuous time Markov chains are used to derive load models and availability models for distributed WfMS \citep{gillmann2000performance}. Also focusing on performance and load management, another approach employs dynamic server assignment where activities are assigned to workflow servers at runtime instead of design time \citep{bauer2000efficient}. 

Much of the work discussed above assumes a central coordination or decision-making authority. Key assumptions are either central coordination with decentralized execution of specific workflow activities, or limited case transfer to a typically homogeneous set of WfMS. The aims are mostly technical, with a focus on infrastructure suitability. Blockchain technology may be seen as another distributed infrastructure technology. However, it differs in key aspects from earlier technology. 
\begin{enumerate}
\item Blockchains replicate all information to all nodes. Selective replication to optimize or minimize communication requirements is eschewed in the blockchain context as it runs counter to independent validation. 
\item As all actors share a consensus view of the workflow state on the blockchain, special decision-making or control nodes are unnecessary, allowing for decentralized control. 
\item Blockchains provide trust by providing a tamper-resistant record. Hence, building control or trust mechanisms on top of the distributed infrastructure is unnecessary. 
\item The proof-of-work consensus method used in most blockchains takes an eventual-consistency approach and accepts latency when establishing consistency, which significantly relaxes the technical requirements on the infrastructure, in terms of performance, security, and reliability.
\end{enumerate}

\subsection{Inter-organizational Workflow Management}

Multiple organizations can collaborate on a single process instance in different ways \citep{van1999process}, such as capacity-sharing, chained execution, subcontracting, case transfer, and loosely-coupled workflows. Blockchain technology can be used to implement all of these collaboration types but may be best suited for the case-transfer collaboration, where all actors share a process definition and each actor performs different activities for a case. Much of the earlier work on distributed workflows (Sec.~\ref{sec:dwfms}), and all of the blockchain-based WfMS discussed below (Sec.~\ref{sec:bbwf}), assumes this type of collaboration.

The public-to-private approach considers a public workflow definition as a contract between participating actors \citep[pg.~141]{van2002inheritance,van2001p2p,van2003inheritance}. Actors can provide private implementations for their parts of a process, as long as these are compatible with the public contract. Compatibility is defined in terms of projection inheritance: The private workflows must inherit the public behavior but may offer specific implementations of this behavior. Public and private workflows are also the foundation for an architecture focusing on flexibility and respect for privacy, where details of local processes need not be publically visible \citep{chebbi2006view}. Inspired by service-oriented architecture (SOA) principles, this approach includes workflow identification and advertisement on a public registry, workflow interconnection governed by contracts (''cooperation policies'') and monitoring using a trusted third party. There is no pre-defined global process model that is partitioned. Instead, the complete process model is dynamically assembled from advertised process interfaces (''public processes'') that describe views on hidden, private (''internal'') processes. In the Crossflow project \citep{grefen2000crossflow}, a process specification forms the contract for interaction among service providers. The technical architecture consists of independent WfMS coordinated by a central contract manager. The contract manager also monitors quality-of-service guarantees. Another use of contracts \citep{weigand2002cross} views them as ''glue to link inter-organizational workflows'' and provides a formal language for business communication. Workflows are managed locally and coordinated among different actors by a central ''contract object'' using messages specified in the contract. Based on P2P networks, \citet{atluri2007decentralized} describe a method to successively partition a complete process model. Each organization receives a process model whose initial activities are assigned to that organization. The organization executes its own activities, then partitions the remainder of the process specification for the successive organizations and passes on those partitions. A central mechanism is only required to initiate each case by identifying the first organization(s), and to accept the final results from the last organization(s). 

Blockchain technology differs from these inter-organizational approaches: 
\begin{enumerate}
\item Each organization acts independently. Executing invalid activities simply leads to transactions that will not be validated by peers and not become part of the consensus blockchain. Trusted third parties for contract monitoring or enforcement are not required. 
\item Blockchain infrastructure makes all transactions publically visible. However, aspects of a workflow may be implemented by each organization privately and the notion of projection inheritance remains useful for this.
\end{enumerate}

\subsection{Blockchain-based Workflow Management}
\label{sec:bbwf}

Blockchain-based workflow execution has only recently received research attention \citep{mendling2018blockchains}. Existing work has focused exclusively on the use of ''smart contracts'' to coordinate workflow activities among participants. A smart contract is a software application that is recorded on the blockchain, ''listens'' for transactions sent to it, and executes application logic upon receipt of a transaction. It can itself generate transactions that can be received by participating organizations.

Driven by a financial institution, a prototype implementation using smart contracts on the Ethereum blockchain offers digital document flow for trading partners in the import/export domain \citep{fridgen2018cross}. The project demonstrates significantly lowered process cost, increased transparency, and increased trust among trading partners. A project in the real-estate domain, also using the Ethereum blockchain and smart contracts, concludes that the lack of a central agency makes it more difficult for regulators to enforce obligations and responsibilities of trading partners \citep{hukkinen2017distributed}. 

The blockchain-based WfMS by \citet{harerdecentralized} uses workflow models as contracts between collaborators. The system allows distributed, versioned modelling of private and public workflows, consensus building on versions to be instantiated, and tracking of instance states on the blockchain. The blockchain provides integrity assurance for models and instance states.

Another implementation of a blockchain-based WfMS uses smart contracts on Ethereum in two ways \citep{weber2016untrusted}. As a choreography monitor, the smart contract on the blockchain merely monitors execution status and validity of workflow messages against a process model. As an active mediator, the smart contract additionally drives the process by sending and receiving messages according to the process model. BPMN models are translated into the Solidity contract language. Peers monitor the blockchain for relevant messages from the contract and create messages to the contract. Each node need not have knowledge of the process definition beyond what is required to produce an appropriate reply to an inbound message. The system checks the acceptability of a response message by running it against a local copy of the contract before publishing it to the blockchain. Transaction cost and latency are recognized as important considerations in the evaluation of the approach. A comparison between the Ethereum blockchain and the Amazon Simple Workflow Service shows that blockchain costs are two orders of magnitude higher than those of a traditional infrastructure \citep{rimba2017comparing}. Recognizing that optimizing the space requirements for smart contracts is important, BPMN models can be translated to Petri Nets, for which minimizing algorithms are available, which are then compiled into smart contracts to achieve up to 25\% reduction in transaction cost while significantly improving the throughput time \citep{garcia2017optimized}. Building on lessons learned from \citet{weber2016untrusted}, Caterpillar is an open-source blockchain-based WfMS \citep{lopez2017caterpillar}. Developed in Node.js and using the Solidity compiler solc and Ethereum client geth, it provides a distributed execution environment for BPMN-based process models. Lorikeet is a similar system \citep{DBLP:journals/insk/CiccioCDGLLMPTW19}, also based on BPMN models that are translated to smart contracts for the Ethereum chain.

Our work differs from prior work in the following aspects:
\begin{enumerate}
\item Our work does not use smart contracts to implement workflow engines for specific workflow models.
\item Our work focuses on the use of standard workflow engines and focuses on the interfaces to the blockchain infrastructure.
\end{enumerate}

\section{Blockchains}
\label{sec:blockchains}

This section describes blockchains that use a proof-of-work consensus mechanism, as implemented in the Bitcoin cryptocurrency and the popular Ethereum blockchain. They are the most common types of blockchains and prior work in blockchain-based WfMS (Sec.~\ref{sec:bbwf}) builds exclusively on such chains.

\begin{figure}[b]
\begin{centering}
\includegraphics[scale=0.5]{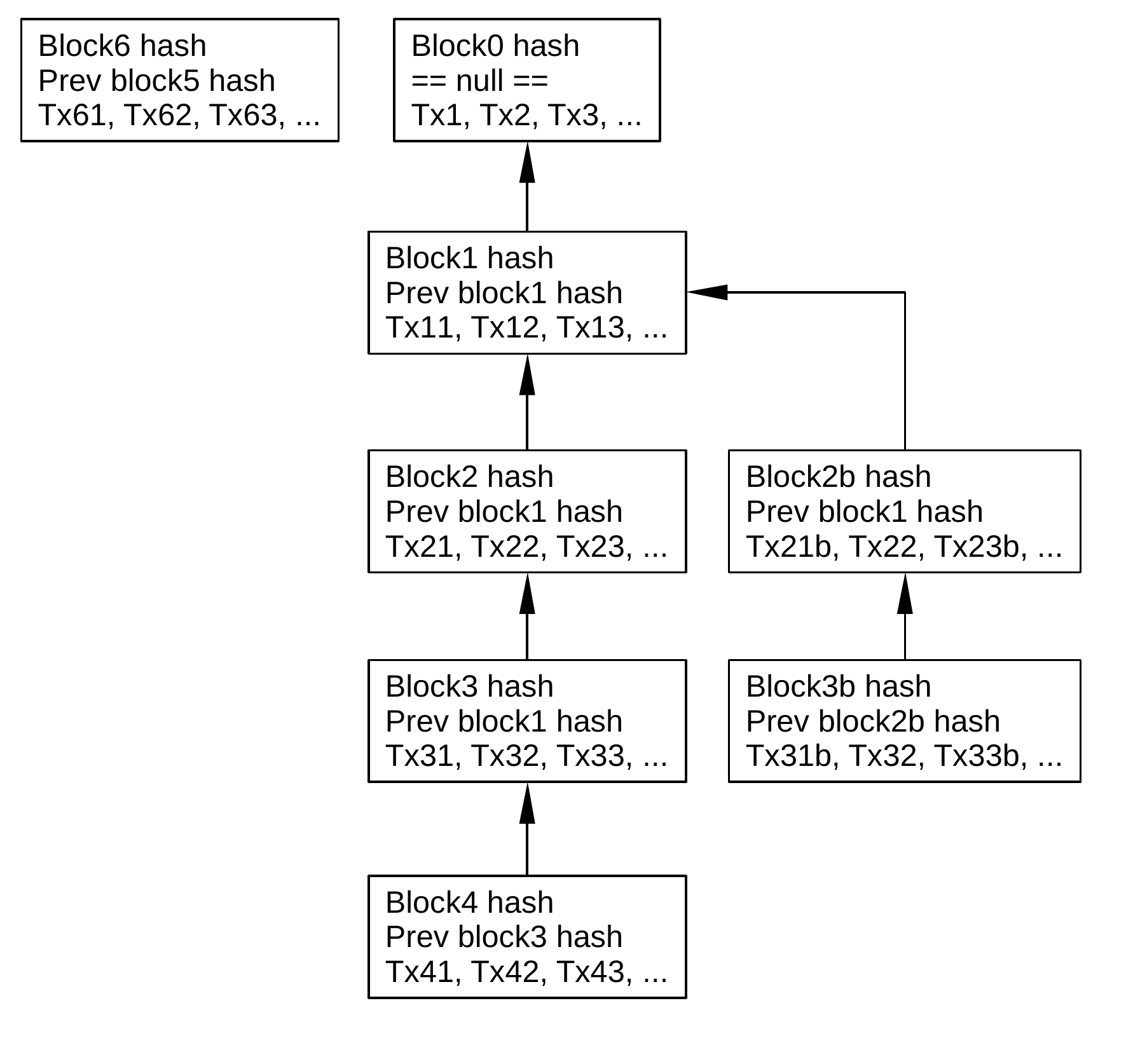}
\caption{Example blockchain with transactions, orphan blocks and side-branch.} 
\label{fig:BlockchainExample}
\end{centering}
\end{figure}

A blockchain consists of blocks of transactions (Fig.~\ref{fig:BlockchainExample}), which can represent any kind of content. Each block also contains the hash of the content of the previous block in the chain. Hence, altering the content of a block requires changing all following blocks in the chain. For example, a change to transaction Tx12 in block 1 in Fig.~\ref{fig:BlockchainExample} results in a different hash for block 1. Hence, block 2's hash needs to be recalculated, and the same for block 3 and block 4.

In a distributed blockchain, new blocks and transactions are distributed among peers. Each peer maintains a pool of transactions to be included in future blocks. New transactions to be added to this pool are \emph{validated}, i.e. it is ensured that they are logically allowed. In the Bitcoin chain this involves ensuring that transaction inputs reference unspent transaction outputs; in the workflow context this may mean that executing a workflow activity is permitted in the current state of a process instance. 

\subsection{Mining}

With sufficient hashing power, it becomes possible for peers to recompute earlier blocks in the chain faster than new blocks are added. Hence, they are able to ''alter history'' as recorded on the chain. To try to prevent such tampering, it must be difficult or expensive for malicious peers to compute block hashes faster than legitimate peers. To make block hashing difficult, the hash is expected to be of a particular form (for example, to have a certain number of leading zeros). This is achieved by adding arbitrary content (a ''nonce'') to the block and repeatedly varying this nonce until a suitable hash is found. This process is known as \emph{proof-of-work mining}. Once a new block has been mined, it is published to all peers.

\subsection{Chain Reorganization}

Depending on network speed, network topology, and other factors, blocks and new transactions arrive at peers in different order and at different times. Hence, each peer may have a different set of blocks and transactions, and hence may also mine different blocks. For example, Fig.~\ref{fig:BlockchainExample} shows blocks 0---4 that reference each other. At the same time, this peer also possesses block 2b, possibly mined by a peer that was in possession of a different set of transactions, and followed by block 3b. Each peer considers the branch with the most mining work (typically its longest branch) as the current \emph{main branch}. Each peer mines new blocks on top of what it considers the head of the current main branch. Side branches occur when different peers mine different blocks based on the same main branch. These may contain different transactions, as in Fig.~\ref{fig:BlockchainExample}, the same transactions in different order, or just a different value for the nonce. Importantly, transactions in side branches are not considered as valid and are not considered when validating new transactions or blocks. 

When a side branch becomes longer than the current main branch, the chain undergoes a \emph{reorganization} as in the following example. In Figure~\ref{fig:BlockchainExample}, assume that block 3b is considered the head of the current main branch and block 3 is the head of a side branch. As block 4 arrives at this peer, block 4 is now the head of the new main branch. As a result, all transactions in blocks 2 and 3, as well as those in block 4, now need to be validated. At the same time, transactions in blocks 2b and 3b that are not in the new main branch are considered invalid and are added back to the transaction pool to be mined again. In our example, these are transactions 21b, 23b, 31b, 33b, etc. as transactions 22 and 32 are also contained in blocks 2 and 3. These will not necessarily be included in later blocks, as they may logically contradict transactions in the main branch.

Blocks may be received for which the predecessor block is missing, such as block 6 in Fig.~\ref{fig:BlockchainExample}. Upon receipt of such an orphan block, a peer will request the predecessor block from other peers until all blocks can be properly linked in the blockchain.

\subsection{Transaction Lifecycle}

A transaction is said to be \emph{submitted} when it is in the transaction pool waiting to be mined. It is called \emph{mined} once it is in the head block of the chain. A transaction is considered \emph{confirmed}, i.e. sufficiently certain to be acted upon, when it has an agreed upon \emph{confirmation depth} from the head of the main branch. For example, in Fig.~\ref{fig:BlockchainExample}, transactions Tx21, Tx22, Tx23, Tx11, Tx12, Tx13, Tx1, Tx2, and Tx3 are considered confirmed at confirmation depth of two, as there are two or more mined successor blocks in the main chain. However, even confirmed transactions can return to the \emph{submitted} state during a chain reorganization. Transactions in orphan blocks or side branches are not considered confirmed.

\section{Architecture, Principles, and Validity}
\label{sec:principles}

The main component of a WfMS is the workflow engine, which interprets the process model, creates and allocates work items for manual execution or executes them using external software applications. The workflow engine maintains state information for each workflow instance (case) as well as workflow relevant information (case data). The workflow engine may be supported by, or include, services for organizational data management and role resolution, for worklist management and user interface, for digital document storage, etc. Designing a blockchain-based WfMS requires choosing where to locate and how to implement the workflow engine and other services. 

Our architecture is for an application-specific blockchain that couples the notion of blockchain transaction validity with the permissibility of a workflow transaction. A workflow transaction may indicate  completion of a work item, launching of case, etc. (cf. Sec.~\ref{sec:prototype}). This is similar to the way in which Bitcoin, also an application-specific blockchain, couples transaction validity to the permissibility of bitcoin spending by examining unspent transaction outputs (UTXO) when validating a transaction. Hence, blockchain nodes require access to a workflow engine to validate workflow transactions. While this requirement admits many different architectural designs, we have opted for the simplest one: each blockchain node has a local workflow engine. While a more general $n:m$ relationship between blockchain nodes and workflow engines may be more flexible in deployments, it does not aid in our aim of identifying specific issues raised by a blockchain infrastructure for workflow management. Our blockchain architecture could be extended to include other types of transactions for other applications, as long as every node has access to validation services for such transactions.

In our architecture, a new workflow transaction originates from one workflow engine. It is passed to the workflow engine's local transaction service, which validates the transaction locally, and, if valid, publishes it on the P2P network. When a node receives a new transaction, it validates the transaction locally before accepting it. When a new block is mined, the miner publishes it on the P2P network. When a new block is received by a node, the block and each of its transactions are validated locally before being accepted. When a block is accepted, each workflow transaction contained in it is passed to the local workflow engine, for it to update its workflow state accordingly. It is important to differentiate between the concepts of \emph{''performing''} a work item in the sense of a human user or an external software application completing a task, and \emph{''executing''} the transaction indicating completion of the work item, i.e. recording its performance and updating the resulting workflow instance state on the blockchain. While work items are \emph{performed} on only one node, the workflow engines on all nodes \emph{execute} the corresponding workflow transaction and therefore maintain a common workflow state.

An alternative design is to use a weaker validity criterion at the blockchain layer, i.e. to accept all properly signed transactions from permitted peers, and let the workflow engine filter out those that are invalid only when they are passed to it for execution. This is the approach used by smart contract based workflow management using generic, application-independent blockchains.

In contrast to our architecture, transaction validity in generic blockchains such as Ethereum, which is frequently used for smart contracts in workflow management, is independent of application-specific semantics of permissibility of the action recorded in the transaction. The validity of a transaction carrying a call to a smart contract method is determined solely by the correct call parameters and sufficient 'gas', not by the application logic of the smart contract itself. The fact that the smart contract accepts a transaction is sufficient for the transaction to be valid. The contract may then decide to simply not update its state when it receives a transaction representing a non-permissible workflow action. Whereas our architecture rejects non-permissible workflow transactions before they are encoded on the blockchain, the smart-contract architecture encodes them on the blockchain, together with the resulting, possibly unchanged, smart contract state. In either architecture, non-permissible workflow transactions are not executed, whether they are rejected prior to inclusion in the blockchain record or after it. 

Our architecture achieves consensus on transaction validity as every node performs independent validation of each transaction that it receives, either for its transaction pool or as block content. A node that attempts to submit a non-permissible workflow transaction will see this transaction rejected by the correctly operating nodes. In a smart contract architecture, transactions that contain non-permissible workflow transactions are accepted but correct nodes will not update the workflow state when the smart contract executes on the local node upon block acceptance. Consider the situation where a faulty (or malicious) node applies a different semantics of permissibility (or a different local smart contract behaviour). The node can maintain the wrong workflow state without this being noticeable, until such time when it submits a workflow transaction based on its faulty state, and which is rejected by other nodes and thus not encoded on the blockchain. Similarly, only when a smart contract call is submitted to the blockchain, but instead of seeing the contract state updated when the transaction is included in the block, the contract state included in the block indicates the dismissal of the workflow action, is the faulty (or malicious) behaviour of the local smart contract replica visible.

The case may arise that competing workflow transactions, say ''A'' and ''B'', both permissible, originate concurrently from different nodes. The fact that some nodes accept A and reject B while others accept B and reject A is correct behaviour as both transactions are valid. This is the same for both types of architectures: On some nodes the smart contract may be called by A first, then the contract must dismiss execution of B. On other nodes, the contract may be called by B first and must dismiss execution of A. The resolution of these conflicts occurs through the ordering consensus, i.e. the mechanism of longest side branches illustrated in Sec.~\ref{sec:blockchains} above and is independent of transaction validity.

In summary, both types of architectures make the same validity guarantees for workflow transactions, and execute permissible workflow transactions on all correctly operating nodes. As long as a majority of peers agrees on what constitutes validity, that set of peers will arrive at a consensus of the blockchain and workflow state.

\section{Prototype Implementations}
\label{sec:prototype}

In this section, we present two implementations of our architecture described in Sec.~\ref{sec:principles}. These research prototypes have allowed us to explore implications of using a blockchain infrastructure for WfMS and to identify possible design choices. For ease of development, they are developed in Java. 

\begin{figure}
\begin{centering}
\includegraphics[width=\linewidth]{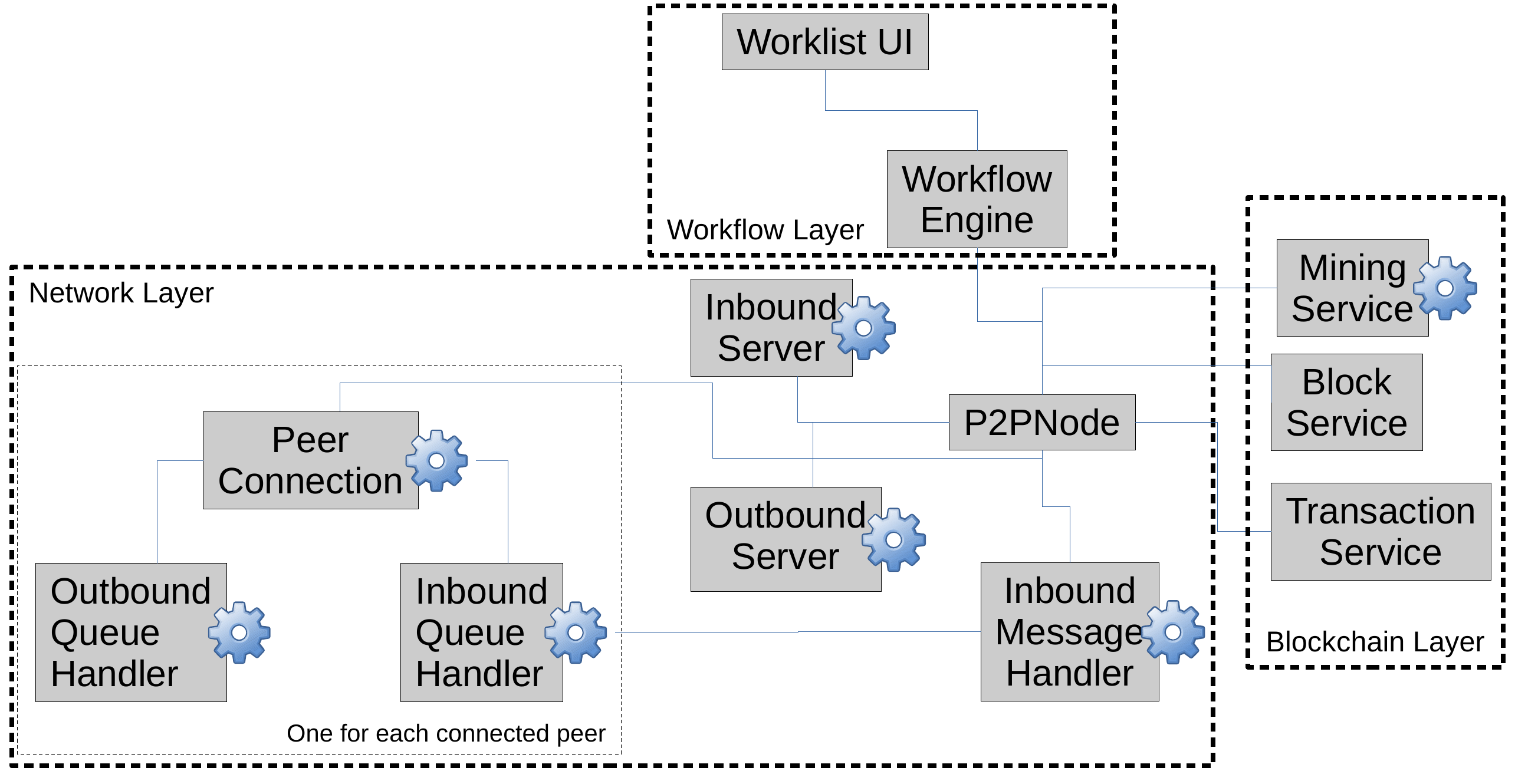}
\caption{Components of the prototype implementation, grouped by layer. Some components run as separate threads as indicated.} 
\label{fig:implementationdiagram}
\end{centering}
\end{figure}

Our architecture has three layers. The network layer forms a private P2P infrastructure with a certificate authority that issues keys to participating actors. To keep our prototype simple, actors are identified by their internet address, rather than their public keys. However, an address resolution layer can easily be added. The P2P layer is implemented using Java sockets and serialization. As indicated in Fig.~\ref{fig:implementationdiagram}, each node has an outbound server that establishes connections to other peers, and an inbound server that accepts and verifies connection requests. Each connection is served by a peer-connection thread, which in turn uses inbound and outbound queue handler threads to receive and send messages. Incoming messages are submitted to the inbound message handler which passes them to the appropriate service. Messages are cryptographically signed and verified upon receipt. Table~\ref{tab:messages} describes the different message types. The P2P protocol is loosely based on that used by the Bitcoin network.

\begin{table}
\begin{tabularx}{\linewidth}{|l|X|} \hline
BlockRequest & Requests a block with a specific hash from a peer \\ \hline
BlockSend & Sends a block to one or more peers \\ \hline
PeersRequest & Requests a list of known peers from a peer \\ \hline
PeersSend & Sends a list of known peers to another peer \\ \hline
TransactionSend & Sends a transaction to other peers \\ \hline
TransactionPoolRequest & Requests the current transaction pool from a peer \\ \hline
TransactionPoolSend & Sends the current transaction pool to a peer \\ \hline
BlockchainRequest & Requests the blockchain, beginning at a certain hash from a peer \\ \hline
BlockChainSend & Sends the blockchain beginning at a particular hash to a peer \\ \hline
\end{tabularx}
\caption{Message types}
\label{tab:messages}
\end{table}

The blockchain layer, comprising the transaction service, block service, and mining service, is implemented on top of the P2P layer. These services appear in all proof-of-work blockchains. The transaction service manages the pool of pending transactions, which are created by the local workflow service or received from the inbound message handler, and are validated upon receipt. The block service receives blocks from the local mining service or from the inbound message handler, validates them, and adds valid blocks to the blockchain. It manages orphan blocks, side chains and blockchain reorganization. For simplicity, our prototypes use fixed-difficulty mining. 

The workflow layer, comprised of the workflow engine and the worklist handler with its user interface, is implemented on top of the blockchain layers. For simplicity, our workflow models are based on plain Petri nets \citep{van1998application}. Each Petri net transition specifies a workflow activity. Each activity is associated with a single participating actor/peer. This is a design decision, as it is also possible to allow multiple participating actors for each activity. The latter is a novel twist on the multiple instance workflow pattern \citep{van2003workflow} as in that case, one, multiple, or all peers/actors may be associated with an activity in order to perform multiple instances of it. The workflow engine keeps track of the Petri net markings for each workflow instance and detects deadlocks and finished cases to remove them from the worklist.

The partitioning of the process to different peers/actors does not form the resource perspective of the workflow; it only signals the workflow engine whether a work item is to be handled on the local node. Our activity specifications allow the process designer to provide further role information, and each node can implement its own role resolution using local organizational information. Similarly, our activity specifications can describe external application calls, which are executed by the workflow engine.

The data perspective is designed as a key--value store. In our prototype, we only admit simple Java types as we implement a GUI for these, but an extension to arbitrary types is readily possible. Each workflow specifies a set of typed variables that can be input and output for activities. These are instantiated for each workflow instance when it is launched. 

When a transition is enabled, the workflow engine on the node to which it is assigned creates a work item and fills its input values from the current values of the workflow instance. The work item is then added to the local worklist. After a work item is performed (manually or through an external application call), output values are written back to the workflow instance. Data constraints can be specified for each activity, which are checked as part of the transaction validation that is performed by the transaction and block services. Upon work item completion, the local workflow engine submits a corresponding workflow transaction to the blockchain.

\subsection{Prototype I: Workflow Actions on the Blockchain} 

\begin{figure}
\begin{centering}
\includegraphics[scale=0.5]{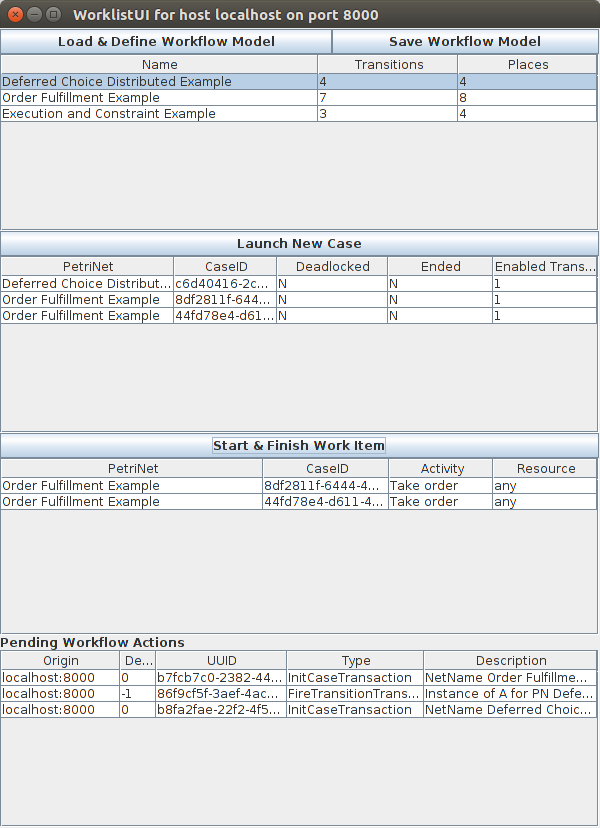}
\caption{Screenshot of prototype I} 
\label{fig:screenshot}
\end{centering}
\end{figure}

In our first prototype the blockchain stores \emph{workflow actions}. We focus on the actions of defining a new workflow model, starting a case and firing a transition. Hence, we define three types of workflow transactions (Table~\ref{tab:transactions1}). Extensions, for example to cancel a case or unload (mark as deprecated) a workflow model, are readily possible. A \emph{ModelUpdateTransaction} defines a new workflow model. For simplicity, we forgo versioning of workflow models and updates to running cases, as this is not relevant to the blockchain infrastructure. An \emph{InitCaseTransaction} launches a new case for a given workflow model. Cases are identified by a universally unique identifier (UUID). A \emph{FireTransitionTransaction} signals that a work item, corresponding to a transition in the workflow model, for a given case has completed, either manually performed or by calling an external application. The work item contains the case ID, the transition name, input data, as well as pre- and post-execution values for output data. Pre-execution values are required for undo ability (cf. Sec.~\ref{sec:issues}). Upon receipt of an \emph{InitCaseTransaction} or a \emph{FireTransitionTransaction}, the workflow engine initializes or updates the data values and Petri net marking in the workflow instance. It then identifies enabled transitions that are assigned to the local node and submits work items for them to the local worklist or executes the specified external application calls. Because the blockchain only stores state changing actions, the workflow engine needs to maintain workflow state, which consists of known workflow specifications, the set of running cases, and the Petri net markings and data values for each case. Figure~\ref{fig:screenshot} shows a screenshot of the prototype, with a list of workflow definitions, running cases, worklisted activities, and pending transactions.

\begin{table}
\begin{tabularx}{\linewidth}{|l|X|} \hline
ModelUpdate & Creates or updates a workflow model definition (Petri net) \\ \hline
InitCase & Starts a case for a given workflow model. Contains model name and unique case ID \\ \hline
FireTransition & Executes a given transition in a workflow model. Contains case ID, transition name, input values, pre- and post-values for outputs \\ \hline
\end{tabularx}
\caption{Transaction types in Prototype I}
\label{tab:transactions1}
\end{table}

\paragraph*{Interface\ }\ In general, any blockchain infrastructure performs three functions. It must accept new transactions for inclusion in the blockchain, it must announce or otherwise provide transactions in new blocks to external applications (as well as invalidated blocks upon a chain reorganization), and it must validate transactions as it accepts them from peers. As we have noted in Sec.~\ref{sec:principles}, because transaction validity includes the permissibility of workflow actions, transaction validation involves the workflow engine.  

These considerations lead to the specific interface that is summarized in Table~\ref{tab:interface1}. The transaction and block services call on the workflow engine to validate transactions. For this, they pass a transaction and the pending transactions in the transaction pool. Validation of a \emph{FireTransitionTransaction} checks that the Petri net transition of the transaction is enabled and that no data constraints are violated. For this, the workflow engine executes the pending transactions for that workflow instance to ensure the Petri net transition remains enabled (i.e. the new transaction is not incompatible with any of the pending ones for that workflow instance). It then undoes the pending transactions in reverse order to restore the current state. Second, the block service passes entire blocks to the workflow engine for execution (to ''do'' them), as described in Sec.~\ref{sec:principles}. Third, during blockchain reorganization (Sec.~\ref{sec:blockchains}), the block service notifies the engine to invalidate blocks of transactions, i.e. to ''undo'' them. Fourth, in the other direction, the workflow engine can get predecessor blocks from the block service, used for retrieving blocks at specified confirmation depths. Fifth, the workflow engine can add new transactions to the transaction pool. With this interface, the workflow engine has knowledge of transactions once they appear in a block at the head of the chain, or when they are locally created. To gain knowledge of all transactions in the transaction pool, an optional interface for the transaction service to notify the workflow engine of new transactions in the pool is implemented. This is not required for the functioning of the WfMS but is useful from the user perspective (Sec.~\ref{sec:issues}).

\begin{table}
\begin{tabularx}{\linewidth}{|l|m{5cm}|X|} \hline
$\rightarrow$ & ValidateTransaction(transaction, pendingTransactions) & Pending transactions are those in the transaction pool \\ \hline
$\rightarrow$ & DoBlock(block) & Announces transactions in the block to the workflow engine for execution \\ \hline
$\rightarrow$ & UndoBlock(block) & Undoes transactions in the block (during blockchain reorganization) \\ \hline
$\leftarrow$ & GetPredecessor(block) & Workflow engine gets predecessor block \\ \hline
$\leftarrow$ & AddTransaction(transaction) & Workflow engine submits new transaction \\ \hline
$\rightarrow$ & {\it AddPendingTransaction(transaction)} & {\it Transaction service notifies engine of pending transaction (optional)} \\ \hline
\end{tabularx}
\caption{Interface between blockchain infrastructure and workflow engine in prototype I ($\rightarrow$ indicates blockchain infrastructure calling workflow engine, $\leftarrow$ indicates reverse direction)}
\label{tab:interface1}
\end{table}

\subsection{Prototype II: Workflow Instance States on the Blockchain} 

An alternative to storing workflow activities on the blockchain is to store complete \emph{workflow instance states}, i.e. data values and Petri net markings. This design does not need separate \emph{InitCaseTransaction} and \emph{FireTransitionTransaction}. They are combined into an \emph{InstanceStateTransaction} that represents a complete workflow instance state (Table~\ref{tab:transactions2}).

This design implies three significant changes. First, because activity execution information is not available in the blockchain, data constraints cannot be specified as post-constraints for each activity, but can only be specified as having to hold for the entire state, i.e. they apply to the global case data (workflow instance state). As in prototype I, constraints are validated when a new transaction or block is received. However, transactions are not validated by executing and then undoing pending transactions. Instead, the workflow engine checks that the marking of the workflow instance state in a new transaction is reachable from the marking of the current workflow instance state as well as the markings of the workflow states in all pending transactions. Second, in contrast to prototype I, the workflow engine does not need not maintain workflow state information as that is readily available by reading the blockchain backwards from the current chain head. This significantly simplifies the workflow engine. Third, the lack of activity execution information means that the user only knows pending future states, but not which pending workflow activities bring about those states. And because it is not always possible to identify which workflow activity brings about a particular state (e.g. multiple enabled transitions that lead to the same state), this information cannot be derived from the current and previous state either.  

\begin{table}
\begin{tabularx}{\linewidth}{|l|X|} \hline
ModelUpdate & Creates or updates a workflow model definition \\ \hline
InstanceState & Specifies the state of a workflow instance \\ \hline
\end{tabularx}
\caption{Transaction types in Prototype II}
\label{tab:transactions2}
\end{table}

\paragraph*{Interface\ }\ The general interface remains the same, as the blockchain infrastructure performs the same three functions of accepting new transactions, validating transactions, and announcing new blocks. However, the specific interface in prototype II (Tab.~\ref{tab:interface2}) differs from that in prototype I (Table~\ref{tab:interface1}). The transaction and block services still call on the workflow engine to validate transactions against the current workflow state. Instead of the ''Do'' and ''Undo'' ability for transactions in blocks in prototype I, the block service simply notifies the workflow engine when a new block is appended to the head of the chain. Hence, the interface is renamed to \emph{UpdateHead}. Instead of executing workflow activities and updating its own workflow state, the workflow engine simply reads the workflow instance states provided in that block's transactions. During blockchain re-organization, instead of providing sets of blocks for undo, the block service notifies the workflow engine that the current blockchain head has been reset to a different block and the workflow engine reads the blockchain backwards from the new head to get the new workflow state. Hence, the interface function is renamed to \emph{ResetHead}. The remainder of the interface is essentially unchanged: In the other direction, the workflow engine can get predecessor blocks, check the confirmation depth of a block, and add transactions to the transaction pool. An optional interface allows the transaction service to notify the engine of pending transactions.

\begin{table}
\begin{tabularx}{\linewidth}{|l|m{5cm}|X|} \hline
$\rightarrow$ & ValidateTransaction(transaction, pendingTransactions)  & Pending transactions are those in the transaction pool \\ \hline
$\rightarrow$ & UpdateHead(block) & New blockchain head is added \\ \hline
$\rightarrow$ & ResetHead(block) & Blockchain head is reset to specified block (during blockchain reorganization) \\ \hline
$\leftarrow$ & GetPredecessor(block) & Workflow engine gets predecessor block \\ \hline
$\leftarrow$ & GetDepth(block) & Workflow engine gets confirmation depth of block \\ \hline
$\leftarrow$ & AddTransaction(transaction) & Workflow engine submits new transaction \\ \hline
$\rightarrow$ & {\it AddPendingTransaction(transaction)} & {\it Transaction service notifies engine of pending transaction (optional)} \\ \hline
\end{tabularx}
\caption{Interface between blockchain infrastructure and workflow engine ($\rightarrow$ indicates blockchain infrastructure calling workflow engine, $\leftarrow$ indicates reverse direction)}
\label{tab:interface2}
\end{table}

We emphasize that while we use Petri net semantics for the workflow models, the interaction and interfaces between blockchain infrastructure and workflow engine are generic and apply to any modelling language and model semantics, including the token-based semantics for BPMN \citep{dijkman2008semantics}. 

\subsection{Comparison}

The two prototypes are motivated by the differences in what is stored in a blockchain transaction: workflow actions or workflow states. We believe that the option to store workflow actions as transactions is the more intuitive approach. On the other hand, it requires a workflow engine that maintains its own state independent of the blockchain. Hence, this option is useful for porting existing workflow engines, as they already possess persistence and transaction management capabilities. Additionally, workflow engines are typically built to process workflow actions, rather than entire states. Storing workflow state on the blockchain simplifies development of new workflow engines, as the blockchain infrastructure can be used not only for distribution but also for persistent storage. The ability to simply read complete workflow states off the blockchain eliminates the need for persistent storage by the workflow engine. Both alternatives require the ability to react to changes in the validation state of blocks, which is arguably easier to implement in the second alternative. Another major difference is that the design in prototype II does not make workflow actions, such as work item completion, explicit. Hence, work item level data constraints cannot be enforced. Explicitly recording workflow actions also allows the blockchain infrastructure to provide meaningful information about pending changes to workflow users, whereas full workflow states may not be informative to users. It is of course possible to store both workflow actions and workflow states on the blockchain to achieve the advantages of both designs, but this leads to high storage requirements and significant redundancy. Finally, storing complete workflow instance states consumes more data on the blockchain than updates only, which may be expensive for public chains but is not an issue for private chains. Table~\ref{tab:prototypedifferences} highlights the differences between the two prototype designs.

 In summary, unless the rapid and relatively easy development or prototyping of an entirely new WfMS is envisioned, as we have done here, the design of prototype I is preferable for adapting or porting existing WfMS onto a blockchain architecture.

\begin{table}
\begin{tabularx}{\linewidth}{|X|X|} \hline
{\bf Prototype I Design} & {\bf Prototype II Design} \\ \hline
Blockchain stores workflow actions & Blockchain stores workflow states \\ \hline
Workflow engine must provide persistence & Workflow engine need not provide persistence \\ \hline
Case-level and work item-level data constraints & Case-level data constraints only \\ \hline
Informative pending transactions & No informative pending transactions \\ \hline
Low data volume & Higher data volume \\ \hline
Suitable for existing workflow engines & Suitable for new workflow engines \\ \hline
\end{tabularx}
\caption{Major differences between the designs of prototypes I and II}
\label{tab:prototypedifferences}
\end{table}

\section{Blockchain Specific Issues in Workflow Management}
\label{sec:issues}

Constructing our research prototypes was useful to identify issues that are presented by blockchain-based WfMS that do not exist in traditional, centralized WfMS, and that have not yet been raised by previous research (Sec.~\ref{sec:bbwf}).

\subsection{Latency}

Proof-of-work-based blockchains introduce latency. At the very least, this is the time between submitting a transaction to the transaction pool and it being mined. A longer latency is introduced when the preferred confirmation depth requires multiple blocks mined on top of a transaction. For example, the Bitcoin community recommends that transactions are not considered as confirmed and acted upon until six or more blocks are added on top of it. Bitcoin mines a new block approximately every 10 minutes. The Ethereum community recommends to assume confirmation at 10 to 15 blocks, with blocks being mined every 13 seconds. Other chains operate at a different pace, but all introduce this type of latency. While latency may not be a problem for slow-moving, long-running workflows where progress is measured in days or weeks, it may be a considerable problem for fast-moving, short workflows that must progress within minutes. In any case, from the user's perspective, workflow activities in a blockchain-based WfMS can remain pending for a significant amount of time, in contrast to traditional WfMS where actions are completed immediately. Users must be aware of this latency and its impact on the workflow. 

\subsection{Validating State versus Visible State}

Workflow transactions go through various stages (Table~\ref{tab:txstages}), from being accepted to the transaction pool, to being mined into a block, and finally to being considered confirmed and therefore actionable. Only the effects of confirmed transactions should be visible to the user, while the effects of all transactions are used when validating new transactions and new blocks. This distinction leads to two different workflow states, which we call the ''validation state'' and the ''visible state''. This distinction is a key difference to traditional WfMS. Understanding the behavior of the WfMS requires the user to have some knowledge about the underlying blockchain infrastructure. In our prototypes we deal with this issue by tracking the status of each pending and mined transaction until it is considered confirmed. Only then is the user's worklist, which reflects the ''visible state'', updated. This can be seen in the bottom part of Fig.~\ref{fig:screenshot}. Providing information about pending and mined but not yet confirmed transactions (i.e. the validation state) allows the user to understand why subsequent activities may not yet be worklisted, or why certain activities cannot be completed, even though they are worklisted in the visible state. We illustrate possible problems due to this discrepancy by examining the sequence and deferred choice workflow patterns \citep{van2003workflow}. 

\begin{table}
\begin{tabularx}{\linewidth}{|m{2.8cm}|m{3cm}|X|} \hline 
{\bf Transaction Stage} & {\bf Validation} & {\bf Note} \\ \hline
Insert into transaction pool & Validate against transactions in chain and transaction pool & Reject invalid transactions but do not remove their work item from worklist; allow user to retry. Remove work items for valid transactions from worklist and report as pending. \\ \hline
Insert to head of chain & Validate against transactions in chain & Reject blocks with invalid transactions. No changes to the worklist. \\ \hline
Reach confirmation depth of chain & & Consider work items for transactions in block as done, and remove them from worklist. Create work items for newly enabled activities. \\ \hline
\end{tabularx}
\caption{Transaction stages, validation points, and workflow actions}
\label{tab:txstages}
\end{table}

\paragraph*{Sequence\ }\ Consider a process where activity B follows activity A and both are assigned to the same node. While the local workflow engine and user knows that activity A has been completed, activity B cannot be worklisted until A is considered confirmed. Or can it? One can imagine a speculative execution of a workflow where the local workflow engine worklists activity B at the risk of having to ''undo'' it at a later stage, should activity A not be accepted by the consensus chain. This is a design choice for the particular WfMS. For this, each node's workflow engine must track the status of its own submitted transactions. In case a transaction is removed from the blockchain or transaction pool, e.g. due to conflicts with other transactions during chain reorganization, it must be undone locally. Speculative execution has not been used to address confirmation latency of blockchain-based WfMS and is an interesting direction for future research.

\paragraph*{Deferred Choice\ }\ Consider a process where either activity A or activity B may be executed. Both activities are worklisted and when A completes, B should be withdrawn, and vice versa. When both activities are assigned to the same node, it may make sense to withdraw B as soon as execution of A is submitted to the transaction pool, as this corresponds to the local user's understanding. However, this may conflict explicit confirmation depth requirement of that user, which may be one or more blocks. From that perspective, B should not yet be withdrawn. Our prototypes implement the latter approach and do not withdraw B from the worklist; completion of A is reported as pending to the user. Of course, completion of B cannot be added as a new transaction. Hence, despite B being worklisted, the user is presented with an error notice upon its completion (not upon its start, as validation can be done only when the engine attempts to add a \emph{FireTransitionTransaction} to the transaction pool). When A and B are allocated to different nodes, the users may not be aware of the deferred choice situation or the execution status of the other activity. In this situation, B should not be withdrawn when A is submitted to the transaction pool: It may be that A does not get mined into the chain, perhaps because the local node mines on what will turn out to be a side branch, or B may be mined into the chain because it arrives at the winning miner before A does. Without understanding the underlying blockchain infrastructure, this behavior will be confusing to a user as it differs greatly from that of a traditional WfMS. 

Confusion can also arise because what is considered confirmed and therefore visible and actionable depends on each node's required confirmation depth of transactions. A user on one node may consider a particular state to permit execution of a following activity, while a user on another node considers that state as insufficiently confirmed. In sequential workflows, this might lead to tensions as one user believes another is delaying the workflow unnecessarily. In a deferred choice situation, this might lead to competitive behavior as one user can always make the choice before the other.

In summary, it is easy to see how the consensus mechanism in proof-of-work blockchains can lead to confusion if users do not have a good understanding of the principles of blockchain infrastructure. To address this, blockchain-based WfMS will require considerable adaptation to user interfaces, as we have begun to show in our prototypes, as well as user training. Earlier research described in Sec.~\ref{sec:bbwf} has not discussed this issue.

\subsection{Confirmed is not Committed: Undo Required}

Section~\ref{sec:blockchains} described how a transaction, even with mined blocks on top of it, may still be invalidated because of chain reorganization; there is no final commit in a proof-of-work blockchain, although invalidation becomes less likely the more confirmations a transaction has. 

Because of this, blockchain-based WfMS require the ability to ''undo'' transactions. As noted above, an ''undo'' ability also enables speculative workflow execution to reduce latency. In prototype I we implement ''undo'' by storing the before-values of all outputs of a work item. An interface of the workflow engine allows the block service to ask for ''undo'' of entire blocks. In prototype II, the undo is performed simply by resetting the blockchain head and reading back the state from the new head. The blockchain is treated as a stack of blocks from which the head block is popped as required. 

\paragraph*{User issues\ }\ Consider again a deferred choice of two activities A and B that are assigned to different nodes. Due to latency, it is possible that activity A is completed and mined into a block, while activity B is also completed and mined into a block. Both blocks are the head of the main branch on their originating nodes, i.e. from the local user's perspectives the work items are completed and the corresponding transactions may even be considered confirmed. However, one of them, assume activity B, will be eventually be ''undone''. Does the user need to be notified of the undo of the transaction so that she can take appropriate action? If so, when and how should the workflow engine notify the user and what information should it provide? Our prototypes add the undone transaction into the list of pending transactions shown to the user, but without highlighting this or raising an alarm for the user. Other implementations could take a more active approach. To take this example further, consider an activity X to be performed prior to A or B on some third node. It may be possible (although unlikely) that X, and therefore \emph{both} A and B are ''undone'' so that both workflow users are presented again with deferred choice of A or B. Both users were convinced that the workflow state includes completion of A or B, respectively, yet both activities are worklisted again. In summary, the eventual consistency approach in proof-of-work blockchains requires users to be made aware of the state of the workflow and each transaction, and be able to understand and make sense of non-intuitive changes to the workflow states. 

\paragraph*{External effects\ }\ In contrast to purely financial transactions such as cryptocurrency transfers (or other virtual transactions, e.g. transfer of ownership), activities in business processes represent considerable human work and other resource consumption. In the best case, undoing workflow transactions wastes this work and the consumed resources. In the worst case, they may not be undoable at all. While financial transactions may be reversible by crediting and debiting appropriate accounts, the performance of workflow activities may bring about substantial and permanent changes of state in the physical world. Applying greater confirmation depth to transactions merely reduces the problem, but does not eliminate it. 

\paragraph*{Compensation\ }\ To address external effects that cannot be undone, the idea of compensating activities or workflow fragments may be useful. Compensation in workflows is a complex issue \citep[e.g.][]{eder1996workflow,grefen2001global,acu2006compensation} but in the blockchain context it raises further questions such as when to worklist compensation activities, whether compensation activities must pre-empt other work items for that case, whether compensation should be done on the validating state, the visible state, or both. For example, consider the blockchain in Fig.~\ref{fig:BlockchainExample} and a user with a confirmation depth requirement of one block. Assume again that block 4 is a new block to be added and block 3b is the head of the main branch. During chain reorganization, transaction 31b in block 3b is invalidated. However, the user has not yet ''seen'' transaction 31b, as her visible state is only up to block 2b. To offer a compensating activity for 31b will be confusing. A compensating activity for the invalidated transaction 21b may be appropriate, but which state, the validating or the visible or both, should be updated with the outputs of this activity? Moreover, the invalidated transactions may refer to activities performed on other nodes, but chain reorganization is a local issue. Should compensating activities be inserted into the original nodes's worklist; yet, that node may not experience the chain reorganization? Should results of compensating activities be confined to the local state only (what about state consensus?) or should they be broadcast on the blockchain (even though some nodes do not undergo a chain reorganization)?

Blockchain WfMS based on smart contracts avoids the ''undo'' requirement as the blockchain will, upon chain reorganization, automatically restore a previous smart contract state because the contract state is encoded on the blockchain itself. However, this does nothing to address the user issues, external effects, and compensation problems highlighted here.

In summary, the ''undo'' required by proof-of-work-based WfMS highlights ambiguities in execution order; it complicates user's understanding of the workflow state and requires user understanding of the underlying blockchain architecture. ''Undo'' may not be possible for some activities or may lead to considerable wasted resources and effort for others. Addressing the problem with compensation may be appealing but raises further difficult questions.

\subsection{Data Dependencies}

Transactions in the transaction pool and transactions within the same block are considered unordered because timestamps in a distributed blockchain are unreliable as there is no central clock. Consider two parallel activities A and B that both write a variable X. The same user first performs A, then B. She therefore expects X to have a certain value. In the absence of control-flow dependencies, both are mined into the same block. As the block and its transactions arrive at nodes, including the originating node, the workflow engines must execute A and B, but in what order? The issue is similarly present when undoing workflow actions. During chain reorganization, the system cannot decide which activity to undo first. Blockchains resolve the ambiguity by executing transactions in block order or by otherwise fixing the order. In any case, the order in which they are executed may not match the order in which they were performed and the value of X may not match the user's expectations. One might argue that even in traditional WfMS the execution order is, in the absence of control-flow dependencies, arbitrary and that, in the presence of data-dependencies, the workflow designer ought to have specified control-flow dependencies. However, in a traditional WfMS the execution order is determined by the user's performance and thus matches expectations.

\section{Discussion and Conclusions}
\label{sec:discussion}

Previous work on blockchain-based WfMS has focused on building smart contracts for specific workflow models on the Ethereum blockchain. In contrast, we have worked with full workflow engines in our own research prototypes, which include case data management, data constraints, and local resource management using standard workflow engine designs. In doing this, we have been able to highlight issues around workflow state visibility, latency, transaction confirmation, and data dependencies that had not yet been considered in WfMS research. 

We emphasize that our research prototypes are not meant for production use; they are research tools for us to identify problems and issues with blockchain-based workflow management in a controlled environment. However, our architectural and interfaces between components, as well as our recommendations below are generalizable beyond the specific prototypes we have presented, and apply to proof-of-work blockchain-based WfMS in general. Specifically, the identified problems and recommendations below apply to smart contract-based implementations as well, as they stem from the properties of proof-of-work chains, not from the specific blockchain architecture investigated here, or our prototype implementations. 

\subsection{Recommendations} 

We have found that blockchain infrastructure introduces peculiarities that are in stark contrast to traditional WfMS. These will require both user interface adaptations as well as user awareness and training. We make the following recommendations.

\paragraph*{User interfaces\ }\ The effects of proof-of-work blockchains cannot be hidden from the user. Hence, rather than trying to hide the infrastructure from the user, we recommend that WfMS design provides full visibility. This includes aspects such as tracking the status of locally and remotely submitted transactions and indicating their confirmation depth, as we have done in our prototypes. User interfaces should also provide informative and constructive feedback and alerts, for example, when users attempt to perform an activity that is incompatible with pending activities, or when a chain reorganization takes place and leads to activities that were assumed to be completed to be worklisted again. 

\paragraph*{User education\ }\ The recommendations for user interfaces are only useful if users are aware of at least the general principles of proof-of-work consensus. Again, hiding the effects of this infrastructure is not possible, so that WfMS users must be educated on transaction states, causes of latency in the system, and the possibility and principles of chain reorganizations. Users must not only be trained on the WfMS, but also on the blockchain, which may pose considerable challenges and detract from the users' actual work. 

\paragraph*{Process designs\ }\ One way to mitigate the effects of blockchain infrastructure is to reduce the number of case transfers between nodes. We recommend that workflow designers consider the sub-contracting pattern\citep{van1999process}, which decomposes activities to sub-workflows of which all activities are assigned to the same node. A hybrid architecture of local WfMS that are joined by a blockchain infrastructure can ensure that only the high-level inter-organizational workflow is affected by the effects of the blockchain infrastructure. Lessons learned from distributed WfMS (Section~\ref{sec:dwfms}) and the use of projection inheritance \citep{van2002inheritance,van2003inheritance} to ensure behavioral correctness apply to such hybrid architectures. Workflow designers should also be increasingly consider compensation activities to be performed when a transaction cannot be undone \citep[e.g.][]{eder1996workflow,grefen2001global,acu2006compensation}. 

\paragraph*{Blockchain Use\ }\ A key motivator for using blockchain-based WfMS is the lack of trust among process participants. While proof-of-work blockchains, in particular Ethereum, are popular with WfMS researchers, other blockchain technologies exist that make the same validity and consensus guarantees but do not exhibit the drawbacks of proof-of-work blockchains. For example, PBFT-based systems (Practical Byzantine Fault Tolerance) systems do not scale well with the number of nodes but offer very low latency and final consensus. Hence, they may be useful for applications with a small number of organizations and and workflows that require low latency and final consensus \citep{vukolic2015quest}. 

\subsection{Smart Contracts versus Application Code}

Prior work on blockchain-based WfMS (Section~\ref{sec:bbwf}) uses smart contracts as workflow engines specific to a particular process model, instead of general workflow engines that can interpret any workflow model, as traditional WfMS do. Smart contracts provide code integrity and visibility/transparency, as the code is part of the blockchain. Additionally, there is no need to call outside the blockchain layer for transaction validation, as we do in our architecture. The disadvantages are the need to re-develop existing application logic. The strong focus on BPMN control flow neglects implementation aspects typically handled by workflow engines such as data management, data transformations, constraints, external services, scripting, decision tables, organizational data management, role resolution, user interfaces, and others. By not porting and reusing traditional workflow engines, implementing these aspects leads to considerable effort for a smart contract architecture, which may be exacerbated due to limitations of the smart contract language instruction set.

In our alternative architecture, the blockchain is treated simply as a trusted infrastructure layer. It serves only to record and share the state of a workflow execution and achieve consensus on the validity of that state. This architecture offers not only the ability to adapt existing workflow engines using simple interfaces (cf. Sec.~\ref{sec:prototype}), avoiding re-implementation effort and relying on proven technologies, but also offers more freedom to implement features that may not be possible in the blockchain execution environment. Implementing application logic off-chain means that developers have access to familiar programming languages, code libraries and development tools. One drawback is that transaction validation must call back to the application logic. Unlike smart contracts, performing validation in off-chain logic places the onus on developers to ensure identical results if nodes use different workflow engines. However, off-chain validation allows developers to develop against a behavioural specification, e.g. BPMN semantics, without specifying the exact algorithms or implementation to be used: The architecture can use a heterogeneous set of workflow engines, each best suited to a particular node's requirements. Table~\ref{tab:arch} lists some advantages and disadvantages of the two architectures.

\begin{table}
\begin{centering}
\begin{tabularx}{\linewidth}{|X|X|} \hline
{\bf Workflow Engine on Blockchain} & {\bf Workflow Engine off Blockchain} \\ \hline
Requires re-development of workflow engine & Can adapt existing workflow engines \\ \hline
Separates workflow engine from external services & Workflow engine remains integrated with external services \\ \hline
Separation of workflow logic from transaction validation & Workflow logic is part of transaction validity; requires call-back to workflow engine for transaction validation \\ \hline
Code integrity \& visibility & Develop against a behavioural specification \\ \hline
Ensures identical behavior for all peers & Behavior must be independently validated by each peer \\ \hline
No design freedom for peers & Allows heterogeneous engine implementations \\ \hline
May be limited by blockchain execution environment & No implementation limitations \\ \hline
May be limited in integrating off-chain components & Few limitations to integrate off-chain components \\ \hline
\end{tabularx}
\label{tab:arch}
\caption{Architectural options for blockchain-based workflow management systems}
\end{centering}
\end{table}

\subsection{Limitations and Future Work}

While we have identified many issues around blockchain-based WfMS and have made recommendations to address them, our work also shows limitations, which we view as avenues for future research. We see both technical as well as empirical research opportunities. 

\paragraph*{Technical research opportunities\ }\ On the technical side, our work has shown potential for further refinement and exploration of architectural options, in particular the following topics:

\begin{itemize}
\item Designing processes to minimize effects of blockchain infrastructure. 
\item Investigating speculative execution of local activities with possible ''undo''. Can speculative execution protocols from other areas in computer science be used and adapted for workflow management?
\item Using compensation activities or compensation workflow fragments to support improved ''undo'' of transactions.
\item Porting existing workflow engines, such as the open-source YAWL system \citep{ter2009modern}, to blockchain infrastructure. This allows more in-depth validation of WfMS as they cover more workflow patterns.
\item Extending the architecture from a $1:1$ relationship between blockchain nodes and workflow engines to an $n:m$ relationship.
\item Implementing BFT-based blockchains for WfMS to identify architectural design issues and implications for WfMS and their users.
\end{itemize}

\paragraph*{Empirical research opportunities\ }\ Neither our work, nor prior work in this area (Sec.~\ref{sec:bbwf}), has advanced beyond prototypes and feasibility studies into production settings yet. Hence, little large scale or in-depth empirical research on user and organizational issues is available. Our recommendations as they affect WfMS users require empirical support. One of the central points is our recommendation to identify effective ways of communicating workflow and transaction state to users, and to make users understand the specific implications of using blockchain infrastructure. Supporting this requires significant observational or experimental work with users and explorations of different WfMS user interface designs, both in a field as well as a laboratory setting. Similarly, the question of whether our architecture can be applied effectively in production settings requires empirical work. Investigating the applicability and feasibility of our architecture needs in-depth case studies into blockchain deployments in inter-organizational settings. 

We believe that application-specific blockchains, like the architecture investigated here, are easier to deploy for stand-alone, ad-hoc applications where the participating organizations have not yet invested into a common, generic blockchain platform. Such situations occur when organizations need to complete workflows in a project setting, instead of a permanent, ongoing basis. One example of such a situation is the permitting of exploration or extraction in the natural resources industry, such as mining and petroleum. Such a process frequently involves multiple stakeholders with different interests and a reasonably structured process of consultation, negotiation, etc. Such a project would prove the ideal site to study the deployment of a blockchain WfMS and its effectiveness in a realistic setting. 

Finally, \citet{mendling2018blockchains} point out the ''people'' factor in adopting blockchain-based WfMS, which they view as an acceptance problem from the enterprise perspective. Through our research, we have identified specific user-focused challenges, such as interface design and user education in blockchain technology, solutions to which will help gain user acceptance. 

To conclude, this paper has described proof-of-concept implementations for an architecture that has not yet seen research attention. Our research shows that workflow engines do not need to be implemented using smart contracts but that traditional workflow engines and the modelling languages they support, can be readily adapted to fit onto a blockchain infrastructure. The interfaces between workflow engines and blockchain infrastructure are simple, and independent of the semantics of the workflow description language. Our work has also highlighted many aspects where blockchain-based WfMS differ from traditional systems. We have discussed implications of these differences and demonstrated how we they can be addressed in our prototype work. In summary, blockchain-based WfMS offer communication, persistence, replication, and trust building in inter-organizational e-Business. 

\bibliographystyle{apalike}

\end{document}